# The point is a trap: Mathematics of Elastic States in Complex Multiscale Systems

Juan G. Diaz Ochoa
Juan.diaz@permediq.de
https://orcid.org/0000-0002-9893-4068
SystMedQ e.V.

## Abstract

Observables in complex biological systems are often reduced in their inherent complexity, resulting in abstract, a priori mathematical representations based on simple concepts such as points in spaces or interconnected nodes in lattices or networks. This reduction is relevant not only for the formulation of multiscale mathematical models and field theories but also for experimental planning and data collection to gain systematic evidence for the nature of such systems, thus extracting robust causal relationships that would lead to better control of the system. However, mathematical abstraction remains challenging and problematic, especially given the tightly intertwined elements between objects and their environment, the system's affordance, and the fact that all these aspects limit fundamental mathematical concepts such as set theory, thus hindering the definition of phase spaces in extreme cases. In this work, I introduce a potential axiomatic representing a nonreducible representation of high-order structures representing entanglement between observables in complex systems and their environment, considering not only their coupling but also their affordance in relation to the overall system and context, and thus intend to formalize concepts already used in complex systems such as Kantian wholes. In light of this axiomatic, I identify states that are rather less elastic with respect to their environment, discuss how low elasticity enables the application of set theory, while high elasticity limits the mathematical representation of complex systems, and present an exemplary Ising model with a network of elastic states with respect to different contexts as a limiting case for mathematical modelling in this context.

## Introduction

The analysis and structural deconstruction of complex systems using abstract mathematical concepts require the exact characterization and grouping of interacting elements into sets (set theory), their representation and complexity reduction with abstract mathematical concepts (e.g., the reduction of interacting proteins as knots in a graph), and the extraction of underlying mechanisms that eventually lead to the discovery of causal relationships or at least causal factors that control the dynamics of the system [1] under the assumption that such causal relationships are essentially grounded in universal physical interactions [2]. These concepts are not only grounded in physics but also in mathematics, systems theory and control theory and are the basis for methods such as reverse engineering of complex systems [3].

Nevertheless, for several complex systems, such as biological systems, it is not possible to deduce evolving phase spaces, thus setting limits to mathematics as well as concepts of causality [4]. It is not only the identification of exact causal relationships and their mathematical encoding but also whether mathematics and corresponding abstractions are enough to represent interactions and causality, encoded as graphs, in any complex system. This concerns aspects in complex systems that go beyond the recognition of higher order representation in complex graphs [5] and that must take into account how the context defines the identity of the elements in the graph.

For instance, some scholars have argued that any object or part of a system can have more than one identity depending on its intended function; for example, a screwdriver can have different uses, more than just turning screws, such as open cans, be used as a tool to remove residue and thus clean objects or even be considered an exposition object in a museum [4]. Therefore, the function/affordance of this object is determined not only by its original design but also by the context and the user. Although objects can be defined as closed elements in a space, their closure cannot be fully defined in terms of its affordance (as a tool, decorative object, status symbol, weapon, etc.), which limits Peano's set theory.

The problem of the identification and characterization of a system and system's elements mentioned above is a challenge in several complex natural systems; some examples are as follows:

- proteins that have different functions depending on the microenvironment [6],
- Organisms that cannot be fully classified (such as facultative mixotrophic organisms, which can be heterotrophs—organisms that consume other organisms—or autotrophs—organisms that produce their own food—and in which symbiotic organisms are able to alter metabolism as well as the innate immune response and cell regulation depending on the context [7]);
- the interactome, i.e., the correct identification of all the mechanisms in organisms (with complex and even changing interacting elements [8]),
- quantum physics and the problem of decoherence [9,10].

Observe that the unstable identity of the system's interacting elements is completely different from that of other concepts in statistical mechanics, such as the concept of a macrocanonical ensemble: any chance of the identity of the system's element implies a change in the ensemble, without implying a particle flow.

All these examples show that the affordance of the element is fundamental for determining the relevance of an element to a given set. While the notion of the Kantian whole essentially refers to evolutionary systems in biology, it is necessary to develop general representations that are not exclusively biological systems and include not only the elements and their corresponding states but also their context. Additionally, this property thus does not limit the formation of meaning in complex systems on the basis of physics but rather goes beyond physics while taking into account the notion of affordability in relation to complexity [11].

In this framework, elements within multiscale systems respond in an elastic way to the context of the system, which is ultimately helpful in identifying specific causal factors that lead to the definition of the behaviour of the system (or phenotype in a biological framework) [1]. Here, multiscale refers to the relationship between the element and its environment: a fine corn of sand is the same size as a cell; however, unlike the corn of sand, a cell can have complex interactions with its environment and across different scales, and depending on the context, it can behave in a more or less complex way. Therefore, the mathematical abstraction of a cell can differ from the mathematical abstraction of the corn of sand: the second can be represented by a point in a space, whereas the first is much more than a point. It is relevant to note that the concept of elasticity is inspired by behavioural economics, as actors are not fully rational in their preferences and react elastically to externalities in their decisions – a concept that has profound implications for how the economy is affected by individual behaviour [12–14].

Therefore, the mathematical definition of an interacting object or a system element cannot be taken for granted [4]. Even the definition of the notion of a point in a set can be too idealized when set theory (based, for instance, on the Zermelo–Frenkel set theory) is applied to the description of real systems. Thus, the concept of the point as a stable abstraction leading to elegant mathematical models may be theoretically feasible, but it is not suitable for developing models that lead to the description of real complex systems.

In addition, compared with real-world physical systems, scaling up and universalizing models pose significant limitations. Consequently, there is a perceptual limit of mathematics, including statistics and data science, as well as probably in physics, in its ability to correctly represent complex systems, the world, and the universe. Evolving systems and the fact that individual interacting elements can show a kind of basal reference to context [4,15]. Such a concept implies that interacting elements and systems require both a relation to their environment and take basal decisions; at the same time, this implies a notion of the existence of such things in themselves [16].

Thus, it is likely that there is a profound problem with elementary axiomatic and a priori knowledge to identify the essential axioms required to define complex systems and for a posteriori knowledge, such as data on changes in the relationship of points to space and time (more on apriorism by DePaul, Michael, and Amelia Hicks [17]). The fundamental mathematical limitation led to a completely different approach: instead of a "Kantian" a priori approach to natural systems, it is perhaps necessary to consider a dialectic between system elements and the entire system and environment.

One way to provide a solution is to avoid the full decoupling of objects and the environment and explicitly consider the intrinsic entanglement between the element and its context. Instead of looking at the conditions under which mathematical descriptions are more realistic [4], several approaches describe such properties, as well as a theory of elastic states and the corresponding effects on system entropy, including the $\boldsymbol{\Phi_S}$ entropy [18]. In all these approaches, however, there is either a lack of formalism or a more precise connection between entropy, $\boldsymbol{\Phi_S}$, and the description of system objects and their context.

In line with the previous discussion, we treat systems with clear, interacting elements that possess a nonreducible and abstract multiscale nature. In the present approach, we deepen the concept of elastic states as an intrinsic coupling between a state and its context (across multiple scales) and explore the idea that the states of the system (as well as the states of the system elements) are not defined a priori but are synthesized continuously. This concept differs from the concept of elastic states in condensed matter, particularly in the topological states of soft materials [19], and represents a further recent development of this fundamental concept in complex systems [15]. The explicit representation of the context within the state is a novel perspective and provides a potentially innovative way to shape mathematical descriptions, which may explain some problems, such as why there are no fundamental theories to describe complex systems or why quantum mechanics cannot be used to explain complex systems [20,21].

The aim of this article is to explore the mathematical perspectives that such a novel definition can offer and how this can be related to novel methods of mathematical modelling of complex systems. We provide a formalism to describe Kantian wholes (CW) as system elements that are intrinsically connected to their environment (i.e., they are not context independent and therefore cannot be isolated or separated from their context). Various forms of element-extended representations such as CW imply the definition of the elastic state (thus providing a stricter definition of a similar concept introduced in previous work [15]).

# Elastic states: the parts and the whole

The mathematical definition of systems usually begins with the idea that the elements of the system (or equivalent parts of the system) have a closure that can be defined mathematically and has invariant properties (in size, space, and time) and that through interactions, the global properties of the system arise from the collective dynamics of such elements. Such mathematical ideas are essentially based on Zermelo–Fraenkel set theory and the axiom of choice [22], which form the basis of every mathematical theory, from analysis to statistics. This axiom assumes that every element in a set is objectively defined within the set and that such a definition is essentially invariant regardless of the boundaries of the set [22].

Such invariance forms the basis of a mathematical truth and the basis of objective definitions. However, the properties of the parts (and the subsequent mathematical definition of a state) are not always invariant, since the elements/parts of the system are embedded in multiple scales and the interaction of these scales cannot be trivially reduced.

## Definitions

Since multiscaling challenges the a priori mathematical definition of an element (and its corresponding states), it is relevant to define what is a context and what is affordance as a relationship between the system element and the context [11]:

***Definition 1***

> Context is defined as the environment where a system is embedded, as well as the environment where a system can be observed; the affordance $\mathcal{F}$ is the dynamical relation between the system and its context.

According to this definition, the simplest context is the reference system (space) in which the system is observed; similarly, the concept of a reservoir in statistical mechanics is an abstract representation of the embedding of a system in an environment or context.

The explicit inclusion of the context in the state is not about the renormalization of parts of the system (for instance, the renormalization of nodes in large networks [23], which requires the explicit consideration of the context that is intrinsically connected to the elements of the system. We explore the possibility of introducing a general definition of states that are entangled with their context. To this end, we present the following definition:

***Definition 2***

A general $object \wedge context$ entanglement, and corresponding states $S$ is defined as an object-context without complete closure and is a system's element that belongs to a group $S \in gKW_i$, where $gKW_i$ (general Kantian Whole) is a group of entangled object-context, and leads to elastic states that noninvariant respect their context.

This definition is general and considers different kinds of entanglements (an object's invariance with respect to its context); Kantian-Wholes are one form of entanglement and thus belong to the general group $gKW_i$.

## Groups of Kantian wholes

The most general definition of a Kantian whole is an abstract representation of the element and its context as simple or complex nodes and refers to them in terms of invariance with respect to the context (see Table 1). For this definition, we essentially assume an entanglement between the element (and corresponding states) and its context (represented as a state entangled with the element's state $\varphi_0$). Different kinds of entanglements thus provide different potential outcomes in statistical systems (see Table 1).

*Table 1 Listing of different system elements and corresponding states with contextual entanglement. Different forms of entanglement imply a degree of invariance with respect to the context, with implications for the mathematical basis of the sampling and the system's mathematical representation.*

| | | |
|---|---|---|
| $S \in gKW_0$ | 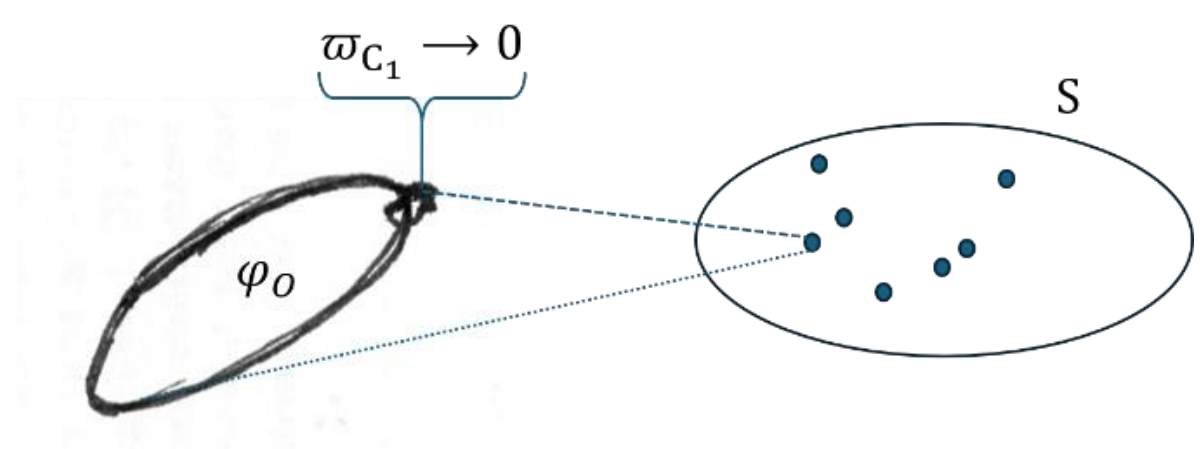 <br>$\varphi = \varphi_0$ | **Context-invariant – neutral contextual embedding and simple affordance:** System's elements are nonentangled to their context $\varpi_{C_i}$. States can be sampled in sets, i.e., $\varphi \epsilon S$. In such case, states (described by $\varphi_O$) can be clearly defined as functions (respect given spaces) in a mathematical way. This case corresponds to classical multiparticle systems. |
| $S \in gKW_1$ | 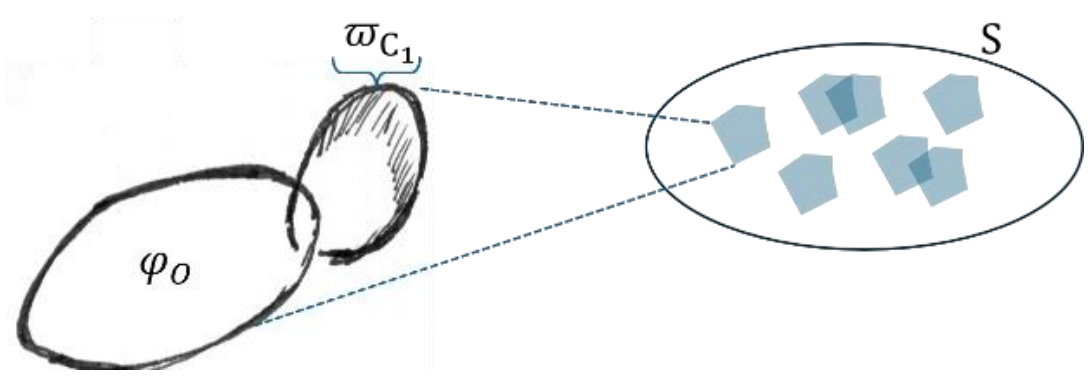 $\langle\varphi\rangle = \langle[\varphi_0 \wedge \varpi_{C_i}]\rangle$ | **Rupture context invariance - contextual embedding and simple affordance:** The context $\varpi_{C_i}$ is defined by the observer and is entangled with the observed states. Such contextual embedding enables the entanglement of states. In such a case, the distribution of the observables in a space can be scanned (represented by $\langle\varphi\rangle \epsilon\, S$). This case corresponds to systems with persistent uncertainty in the system's observation. |
| $S \in [gKW_2]_i$ | 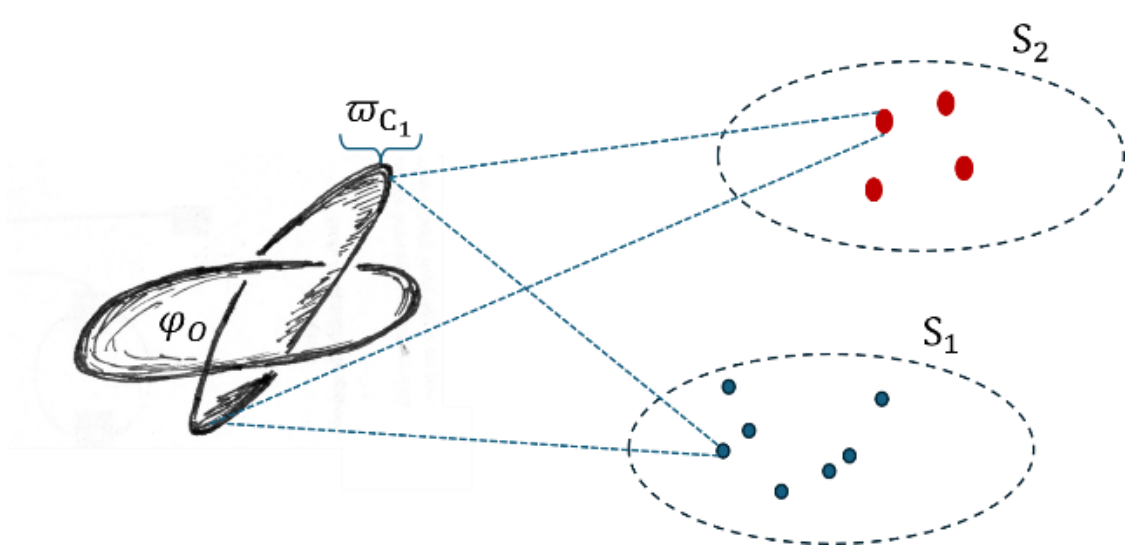 $\left\langle \begin{matrix} \xi \\ \updownarrow \\ \varphi \end{matrix} \right\rangle \triangleq \left\langle \begin{matrix} \varphi_0 \wedge \varpi_{C_2} \\ \updownarrow \\ \varphi_0 \wedge \varpi_{C_1} \end{matrix} \right\rangle$ | **Mult scaling rupture of context invariance - contextual embedding and affordance:** The entanglement between the observed state and its context $\varpi_{C_i}$ is not trivial. It also implies that context changes can lead to abrupt and nonlinear (and eventually not mathematically reducible) changes in individual states. |

All these different types of elements that lead to different states allow for a nuanced definition of complex systems. Moreover, the context defines control factors that limit local states.

The characterization of the effect of each state on the system's performance is described by $\boldsymbol{\Phi}_S^k$, which describes the degree of system variability and autonomy [24]; i.e., for each element-object $k$ that belongs to $gKW_i$, both the internal structural changes in the system (variability) and adaptive autonomy (related to affordance) depending on changes in the system's context are assessed [24]. This entropy is a vector-like measurement and is defined as follows:

$$\boldsymbol{\Phi}_S^k = [\Phi^k{}_{GP}, S(\Gamma^k)] \tag{1}$$

where $\Phi^k{}_{GP}$ is the relative distance between persistent patterns computed from time periods extracted from the trajectory's time series [24] and is defined as follows:

$$\Phi^k{}_{GP} = min\, D_{KL}[L[w^k{}_\tau]||L[w^k{}_{\tau'}]] \tag{2}$$

where $D_{KL}$ is the Kullback–Leibler distance of the persistence patterns $L[w^k{}_\tau]$ of time periods extracted from the trajectory $\Gamma$ (defined as a time series) of the system $k$. This measurement essentially is an estimation of potential causal structures leading to the trajectory $\Gamma$.

The second complexity measure accounts for the system's variability, i.e., how strong is the variation (relative difference) between different systems (organisms) [18,24] and is defined as follows:

$$S^k = \sum_l \Delta_{kl} \cdot \log(\Delta_{kl}), \tag{3}$$

where $\Delta_{kl}$ is the relative difference between the persistent patterns in the system's trajectories $\Gamma$ of system $k$ and those of the other reference system $l$.

In summary, $\mathbf{\Phi}_S^k$ provides an estimation of inherent causal structures and system variability for each element/object in the system. The total $\mathbf{\Phi}_{\boldsymbol{S}}$ entropy, computed over the whole system's elements, provides an estimation of system variability and the persistence of inherent causal structures. Furthermore, this entropy leads to a notion of observability when $\mathbf{\Phi}_{\boldsymbol{S}} \leq [I_{GS}, I_S] = \boldsymbol{I_S}$, where $I$ is the integrated information of the join probability computed from the patterns of the sampled trajectories [24]. The original definition is valid for time series in systems far from equilibrium, and the implicit measurement of distances provides information about changes in the context of the system.

In the case of equilibrium, sampling of all system states leads to a system trajectory that follows not only the optimum in an energy landscape but also the affordability of the system; therefore, variability and autonomy are not relevant, and only the quantification of the distributions in equilibrium is relevant. In that case, $\mathbf{\Phi}_{\boldsymbol{S}}$ entropy is essentially the distance between different trajectories in equilibrium,

$$\mathbf{\Phi}_{\boldsymbol{S}} = D_{KL}\left[\Gamma_{C_i}, \Gamma_{C_j}\right] = I, \tag{4}$$

where $D_{KL}$ is the Kullback–Leiber distance: assuming that the trajectories sample the states $\Gamma_{C_i} = (\varphi_1, \dots, \varphi_n)$ and $\Gamma_{C_i} = (\varphi'_1, \dots, \varphi_n')$, defined respect two different contexts $C_i$ and $C_j$, then $D_{KL}\left[\Gamma_{C_i}, \Gamma_{C_j}\right] = \sum_{l=1}^{n} \varphi_l \ \log\left(\frac{\varphi_l}{\varphi'_l}\right)$. Since we are dealing with systems in equilibrium, the difference in trajectories related to the system's context provides information about the system's structure and affordance, i.e., $\mathbf{\Phi}_{\boldsymbol{S}} = I_S \triangleq \mathcal{F}$.

From the taxonomy introduced in Table 1 and the $\mathbf{\Phi}_{\boldsymbol{S}}$ entropy, the following three main groups can be defined.

## Noncontextual: $\boldsymbol{S} \in \boldsymbol{gKW}_0$ – classical mechanics and the concept of particle

In noncontextual $gKW_0$ groups, the parts/elements of the system can be isolated and defined neutrally and absolutely; i.e., the elements of the system are invariant in space, scale, and time. In other words, such systems are context invariant: regardless of a change in context, observations are invariant or can be extrapolated (transformed) into the new context, for example, by transforming reference systems (relativity principles).

In complex systems, elements interacting inside a network structure have a control function, which can be defined in physical terms such as, for instance, a Hamilton function of the form $\mathcal{H}|_{C_1} = -\sum_{i,j}\left[g_{ij}\varphi_i\varphi_j\right]|_{C_0} - \alpha.\varphi_i$, which determines the system's state through its optimization,

where $g_{ij}$ is the coupling factor of the states $\varphi_i$ and where $\alpha$ is an external factor (field) affecting the system. The system's "affordance" is essentially determined by the optimization of the control function $\mathcal{H}$, which is invariant with respect to different contexts, i.e.,

$$Opt(\mathcal{H}|_{C_1}) = Opt(\mathcal{H}|_{C_2}) = \cdots = Opt(\mathcal{H}|_{C_k}) = \mathcal{F}. \quad (5)$$

Correspondingly,

$$S \in gKW_0 \Leftrightarrow \boldsymbol{\Phi}_S \to 0. \quad (6)$$

In general, this case is rooted in a classical statistical system with well-defined states, leading to collective dynamics based on fundaments of classical mechanics.

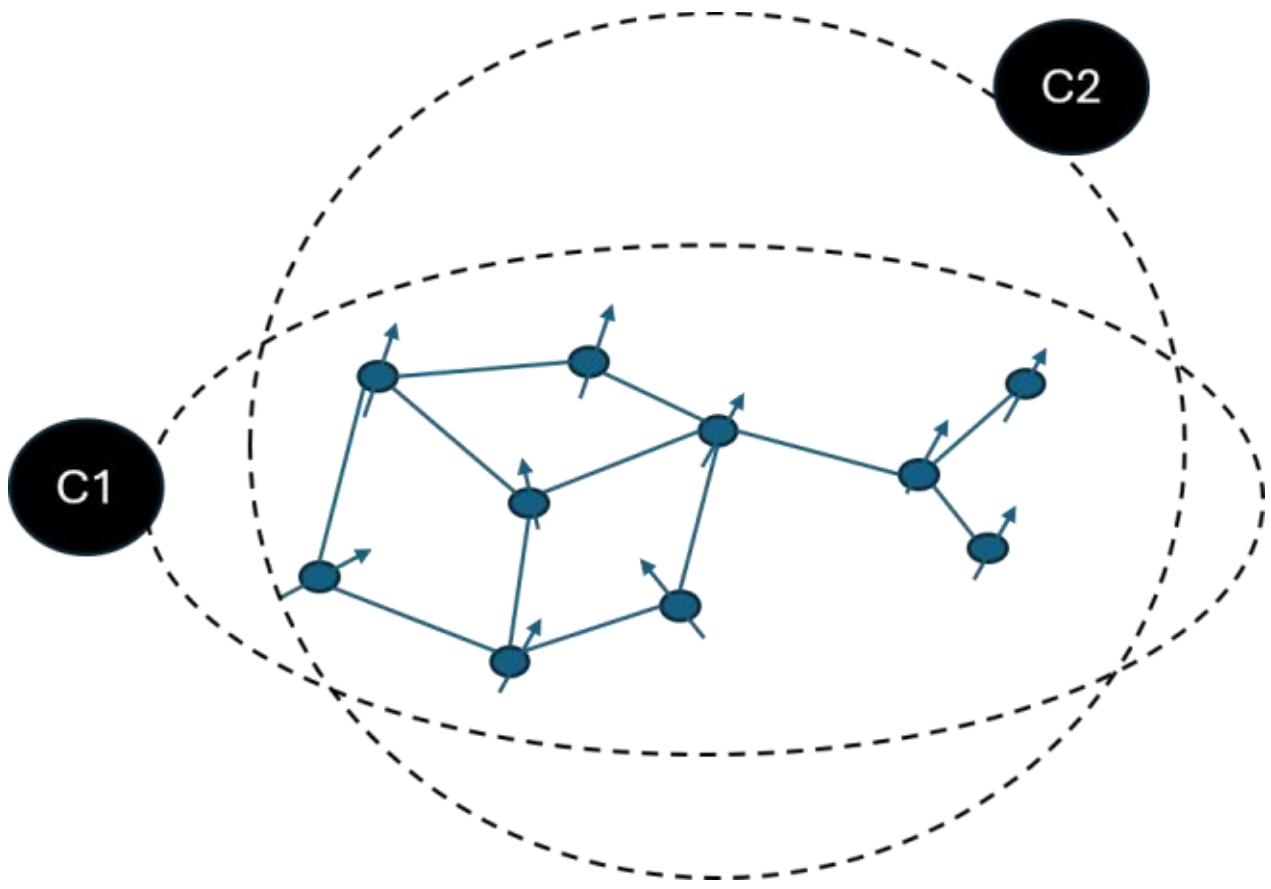

*Figure 1 noncontextual $S \in gKW_0$ implies a system's invariance with respect to the context. The system's elements and interactions are context invariant, even when dynamic interactions and particle flow are considered. This category corresponds to classical mechanistic and causal modelling (such as Ising, Boolean and differential equation modelling).*

Such systems can be described using mechanistic or stochastic methods such as Monte Carlo methods; in the case of complex systems, they can usually be represented as graphs (networks) with interlinked elements (determined by the linking factor $g_{ij}$) with states defined by functions $\varphi_i = \varphi_i(t)$ that evolve in time (an exemplary network, context invariant, is represented in Figure 1). This means that such systems are context independent and that the overall dynamics depend on the system's parameters and boundary conditions.

## Contextual: $\boldsymbol{S} \in \boldsymbol{gKW_1}$ – concept of states with a distribution and uncertainty

In the next group, the context influences the distribution of the system elements (and thus its individual observability). Coupling with the context does not allow for the definition of exact states; instead, the distribution of all possible states must be sampled. By explicitly including the context, we provide an additional element in the observed state that encodes the context and is not always accessible to the observer (for example, in the context of context 1). Therefore, the observer is equivalent to a context and influences the observed objects. Quantum mechanics can probably be described as a system belonging to a $S \in gKW_1$ group with elements with simple coupling to their context.

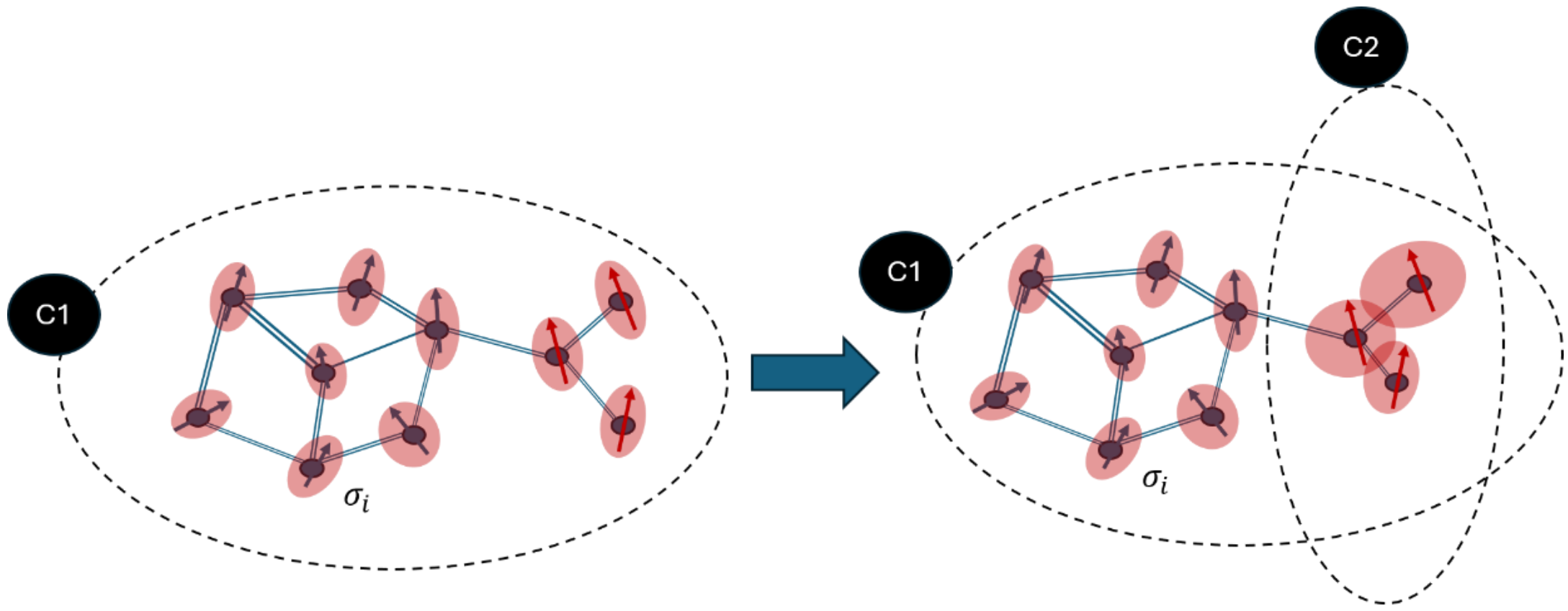

*Figure 2 Group $S \in gKW_1$, where changes in the context perturb the distributions. Such systems are essentially ergodic, such that they deliver similar functionality in equilibrium. Despite perturbations in single states, the control function can be sampled, allowing a statistical description of the system.*

Coupling to context implies that there is no necessary information about which context is currently coupled to each state. The central concept here is the explicit consideration of a state based on an entangled context, which leads to uncertainty that is not trivially accessible and implies that objects and their corresponding states cannot be reduced to points but rather to operators and distributions [25]. This aspect is outlined in Figure 2, where states are represented by distributions (symbolically by transparent ovals) that change in their distribution (and can even be entangled) when a new context is considered.

Mathematically, this implies that states are affected by operators, i.e., $\widehat{\mathcal{H}}_C$; in the case in which field representations are needed, the control/energy function is represented as a function of operators, i.e., $\widehat{\mathcal{H}} = \omega\left(\boldsymbol{\gamma}^{+}\boldsymbol{\gamma} + \frac{1}{2}\right)_{C_0}$, where $\boldsymbol{\gamma}$ are operators containing information about the momentum and position of the states. Thus, when context is entangled with the observed elements, they are essentially represented in terms of operators and not point particles in a space with an exact relation between momentum and position [26]; simultaneously, single states cannot be represented as points but as distributions $\langle\varphi\rangle$ (Table 1).

In this framework, while the elements are entangled with the environment, the operators (the specific way to observe the system) are invariant with respect to the context once the observation has been defined; i.e.,

$$Opt\langle\widehat{\mathcal{H}}|_{[C_1]}\rangle = Opt\,\langle\widehat{\mathcal{H}}|_{[C_2]}\rangle = \cdots Opt\,\langle\widehat{\mathcal{H}}|_{[C_i]}\rangle = \mathcal{F}, \tag{7}$$

where the optimal operator (for instance, associated with a minimal action) determines the system's affordance $\mathcal{F}$. Such invariance is the reason why it is possible to define mechanical principles, despite the uncertainty in the system's elements. Similarly, single states cannot be fully characterized, but the whole system can, under specific experimental conditions, be described in a mathematical way, i.e.,

$$S \in gKW_1 \Leftrightarrow \boldsymbol{\Phi}_S \rightarrow 0. \tag{8}$$

Note that in this expression, we do not focus on causality but focus much more on the variability of the system. The apparent influence of the observer is thus automatically contained in the connection of the element with its context and does not presuppose the emergence but the accessibility to states that are more or less connected to their context (and not only to other states); however, this coupling does not lead to nonlinear changes in field interactions or to significant system variability.

This approach can help explain why decoherence exists in quantum systems: depending on the observation scale, we treat systems that ultimately require explicit consideration of context, which implies that any interaction with this state type implies at the same time an interaction with its context within a group $S \in gKW_1$; otherwise, in context-agnostic systems, we return to a group $S \in gKW_0$. In the case of simple coupling, a sample of $S \in gKW_1$ is carried out, which leads to distributions with stable physical constraints [9].

## Contextual: $[\boldsymbol{S} \in \boldsymbol{gKW}_2]_{\boldsymbol{i}}$ – context and affordance in complex systems

In the following group, we address elements defined across several scales, owing to more complex states as well as more complex entanglement with the context. In such groups, the system's elements have a nonreducible multiscale character and have a more complex relation to the system's affordance $\mathcal{F}$, either preserving the symmetry of $\mathcal{F}$ or seeking different entanglements with their context (also looking for different affordances).

These are therefore systems that are difficult to represent mathematically. We can identify from this definition two main groups:

### Contextual: $[\boldsymbol{S} \in \boldsymbol{gKW}_2]_1$- Case 1

In this group of system elements, the system changes its identity and thus its overall structure to maintain a function within a context when the system is disrupted (for example, by shutting down a system element). For instance, *prolonged abnormal forces (e.g., hypertension, fibrosis, and protein variants) are 'sensed' by and propagated through biological structures in cardiac muscle, ultimately inducing gene expression programs that mediate the remodelling of cell and tissue structure* [27].

Therefore, the system's structure changes and adapts to keep affordance invariant (see sketch of this concept in Figure 3).

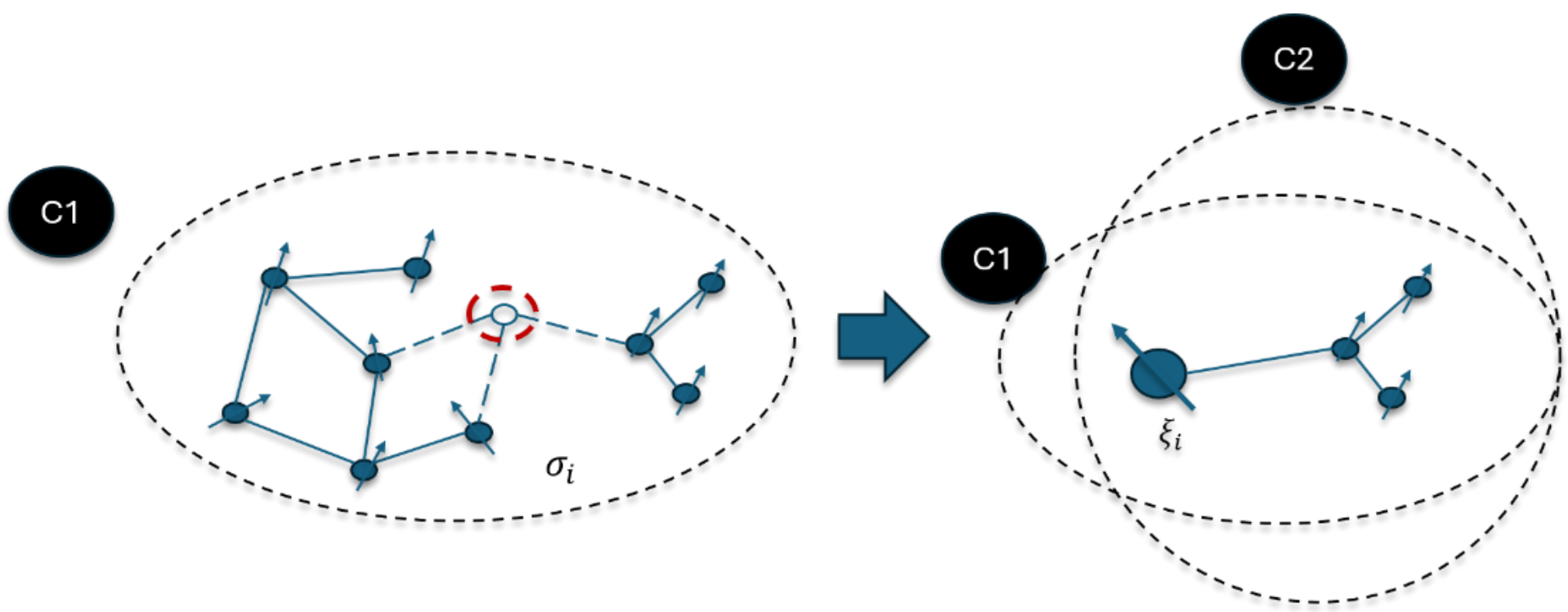

*Figure 3 Diagram of a system with the perturbation of internal states preserving context-affordance invariance. Such a system belongs to an $[\boldsymbol{S} \in \boldsymbol{gKW}_2]_1$ group.*

In such cases, the simplest elastic state implies a change in the element's connectivity or node grouping to preserve the affordance $\mathcal{F}$:

$$Opt\,\langle\mathcal{H}(\sigma)|_{[C_1]}\rangle = Opt\,\langle\mathcal{H}'(\zeta)|_{[C_1,C_2]}\rangle \triangleq \mathcal{F}. \tag{9}$$

The most complex situation implies both a change in connectivity and a change in the identity of the element (state or reaction type of the element). In both cases, the causal factor (related to the functionality of the system) is invariant, which allows the definition of a space with clear constraints. The corresponding entropy is

$$S \in gKW_{i>0} \Leftrightarrow \boldsymbol{\Phi}_S > 0, \tag{10}$$

This means that there is both system variability and autonomy, but the constraint of the invariant affordance still maintains low entropy

## Contextual: $[\boldsymbol{S} \in \boldsymbol{gKW}_2]_2$ - Case 2

In the second case, we look at changes in the context and thus in the affordance of the system, such as facultative mixotrophic organisms, which can be either heterotrophs or autotrophs, in which symbiotic organisms can alter metabolism as well as the innate immune response and cell regulation, depending on the context [7].

Taking into account the entanglement between the interacting objects and their context, this implies a constant recalibration of the system to the different contexts (see diagram in Figure 4).

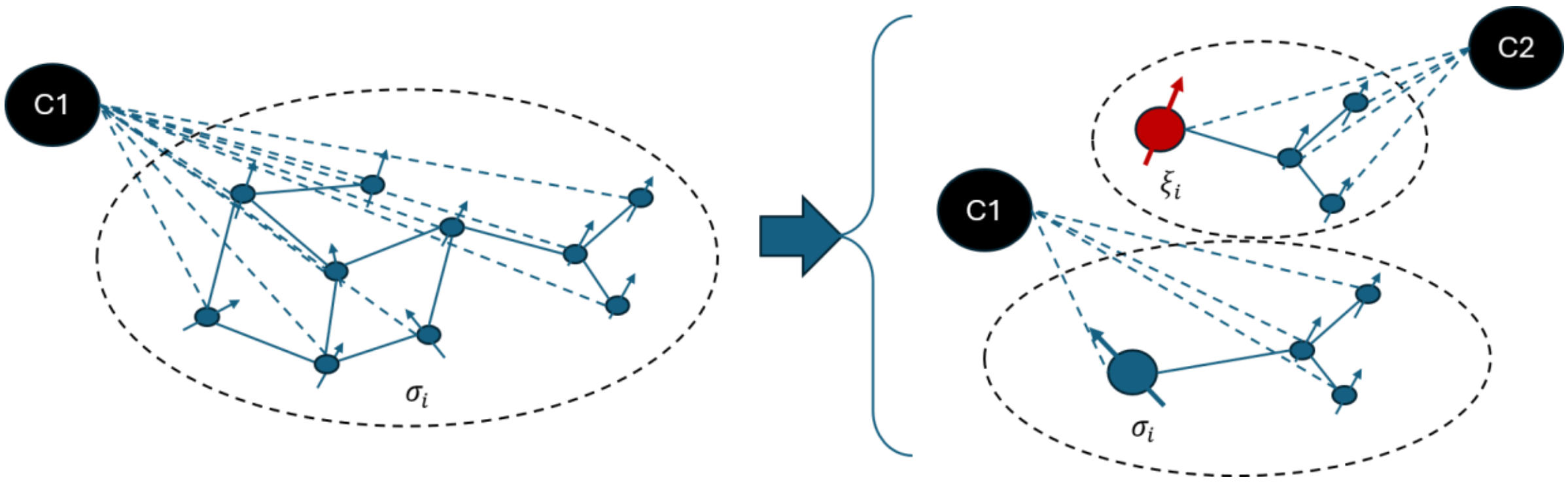

*Figure 4 Diagram of a system with changes in its structure and single elements due to changes in context. Such a system belongs to an $[\boldsymbol{S} \in \boldsymbol{gKW}_2]_2$ group.*

In this case, the affordance $\mathcal{F}$ is noninvariant; the constant readjustment to the context led to a constant calibration of the system's affordance (which in some cases can be related to teleonomy).

The group $[S \in gKW_2]_2$ sets a clear limit to any mathematical definition and implies a limit in the possibility of establishing system equations, with abrupt changes in the mathematical forms, i.e.,

$$\begin{matrix} \mathcal{F}' & \triangleq & Opt(\mathcal{H}(\xi)) \\ \updownarrow & & \updownarrow \\ \mathcal{F} & \triangleq & Opt(\mathcal{H}(\sigma)) \end{matrix}. \tag{11}$$

Such extreme system object–context entanglement implies high variability as well as system autonomy, implying a divergent $\boldsymbol{\Phi}_S$ entropy. Accordingly,

$$S \in gKW_{i>0} \Leftrightarrow \boldsymbol{\Phi}_{\boldsymbol{S}} \gg \boldsymbol{0} \tag{12}$$

This relationship provides us with information about the universe's mathematical reducibility, as well as its ability to create and continuously evolve.

## Example: modified Ising with elastic states

The kinds of states defined as $gKW_i$, introduced in the previous section, provide the basis for the definition of systems that can be represented in a mathematical way but with clear constraints: while states of classes $S \in gKW_0$ and $S \in gKW_1$ are associated with systems that can be observed (and can thus deliver coherent data) as well as mathematically represented (using structural models encoding fundamental interactions across the system's elements), $S \in gKW_2$ pose difficult constraints for this mathematical representation.

In particular, the class $S \in gKW_1$ belongs to a class of systems in which the states can be represented by distributions, while the system generally remains context independent, which facilitates the mathematical description of such systems. In the case of quantum mechanics, the observer perturbs the state without changing the fundamental energy condition. On the other hand, the class of states $[S \in gKW_2]_i$ is more challenging, since there is in principle no objective, complete and unique mathematical definition of the system. This implies that such systems cannot deliver reliable data (when observed) and are in principle intractable from a mathematical perspective.

To illustrate the dynamics of a general group $[S \in gKW_2]_{k\epsilon[1,2]}$, we present a modified Ising model as a relatively simple way to illustrate how such states might work; Ising models are used to represent the magnetization properties of ferromagnetic materials but are also used (and misused) to represent other kinds of systems, from biology to sociology and economics [28]. The advantage of such an implementation is the ability to understand phenomena across scales, such as phase transitions.

An Ising model consists of coupled elements in a lattice or a network that can adopt a binary state $\sigma_{\mathrm{i}}$, i.e., $\sigma_{\mathrm{i}} = -1$ or $\sigma_{\mathrm{i}} = 1$. The coupled states can have different binary states as well, and the sum of the coupled states defines the system's control function (which in physics is designed as the system's Hamiltonian). In the basic implementation, the Hamiltonian $H \in gKW_0$ is defined as follows [29]:

$$\mathcal{H}(\varphi) = -\sum_{i,j} g_{ij}\, \varphi_{\mathrm{i}}\varphi_j - \sum_i \alpha \cdot \varphi_{\mathrm{i}} \tag{13}$$

where $g_{ij}$ is the coupling of the spin states $\sigma_{\mathrm{I}}$ and represents the weight (strength) of the edge connection; in addition, this coupling encodes the structure of the system (as a grid or network with a given architecture). The α parameter represents an external field parameter and gives the nodes a direction preference regardless of their neighbors. The distribution probability of the states is given by $P(\sigma) = \frac{e^{-\beta\mathcal{H}(\varphi)}}{Z}$, where $Z$ is the partition function defined as the sum over all the system's configurations $Z = \sum e^{-\beta\mathcal{H}}$. Thus, the local states influence the total energy-control function, whereas the optimization-minimization of the control function leads to the dynamics and final equilibrium of local states $\sigma_{\mathrm{i}}$. The initial Ising model has 50 nodes distributed in a small-world network. We can obtain small-world networks starting with a ring lattice followed by randomly rewiring edges. If this process is continued, we obtain a random graph. Between the ring lattice and random graph, we find the small-world network. It has local clustering of the ring lattice and low average distances from the random graph. Obtaining small-world properties as a mix of a ring lattice and random graph is referred to as the Watts–Strogatz model; each state is actualized on the basis of a Metropolis algorithm [30].

Additionally, we implemented the class $[S \in gKW_2]_2$, which is mathematically tractable (the causal function remains invariant and maps to the system's function; $[S \in gKW_2]_1$ is partially tractable when the change in context is embedded in the model, for instance, by changes in connectivity that coevolve with changes in the control function [31]).

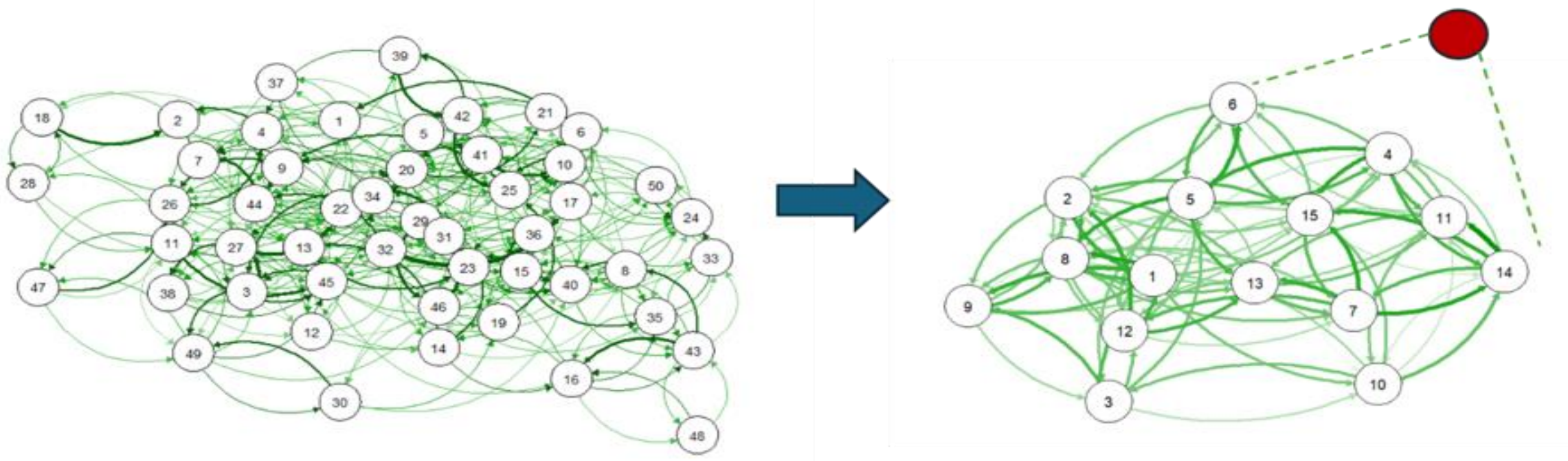

*Figure 5 Implemented model according to a $[S \in gKW_2]_2$ group*

In the implemented model, we have a relevant difference from a classic Ising model: the network topology and node distribution change depending on the context (in this case, depending on the parameter $\beta$); thus, according to the group $[S \in gKW_2]_1$, the Hamiltonian is redefined depending on the context. Consequently, some states "freeze" and start behaving like an external field (see Figure 5); thus, the initial model is a small-world network with 50 nodes, whereas for $\beta < \beta_T$ ($\beta_T$ a threshold $\beta$ value), the node transformation leads to a smaller network (random transformation, with a network size of 10% to 70% of the original network size). According to equation 7,

$$\begin{array}{ccc} \mathcal{F}' & \triangleq & Opt(-\sum_{i,j} \; g'_{ij}\, \xi_i \xi_j - \sum_i \alpha'(\xi) \cdot \xi_i) \\ \updownarrow & & \updownarrow \\ \mathcal{F} & \triangleq & Opt(-\sum_{i,j} \; g_{ij}\, \varphi_i \varphi_j - \sum_i \alpha \cdot \varphi_i) \end{array}, \qquad (14)$$

where $\xi_i$ are still binary states that can adopt 1 and -1 values; however, the frozen part $\alpha'(\xi)$ implies that the system adopts a different affordance $\mathcal{F}$ (see Figure 5). Thus, we have two scenarios: a phase space, embedded in a monotonic variety in a simple context, and a phase space in a fractional variety.

In the first case, the transition from a disordered state to an ordered state governed by alignment occurs suddenly at a critical point as we increase β (pitchfork bifurcation). This is equivalent to a symmetry break: the disordered state is embedded in the flat manifold of the phase space, which is folded when the symmetry breaks. In the disordered phase, we have linear control over the state of the system by the external field. We lose this control in the ordered alignment phase. Here, perturbations in the outfield have little effect, and the system becomes resistant to them. Additionally, alignment induces two tipping points, between which the state of the system depends on its history. This also means that once a tipping point has been exceeded, the change is difficult to reverse [29]. In accordance with the theory introduced in the previous section, this manifold, and its corresponding folding, belongs to a single context $C_1$ and has a low $\Phi_S$ entropy. This effect is shown in Figure 6 A.

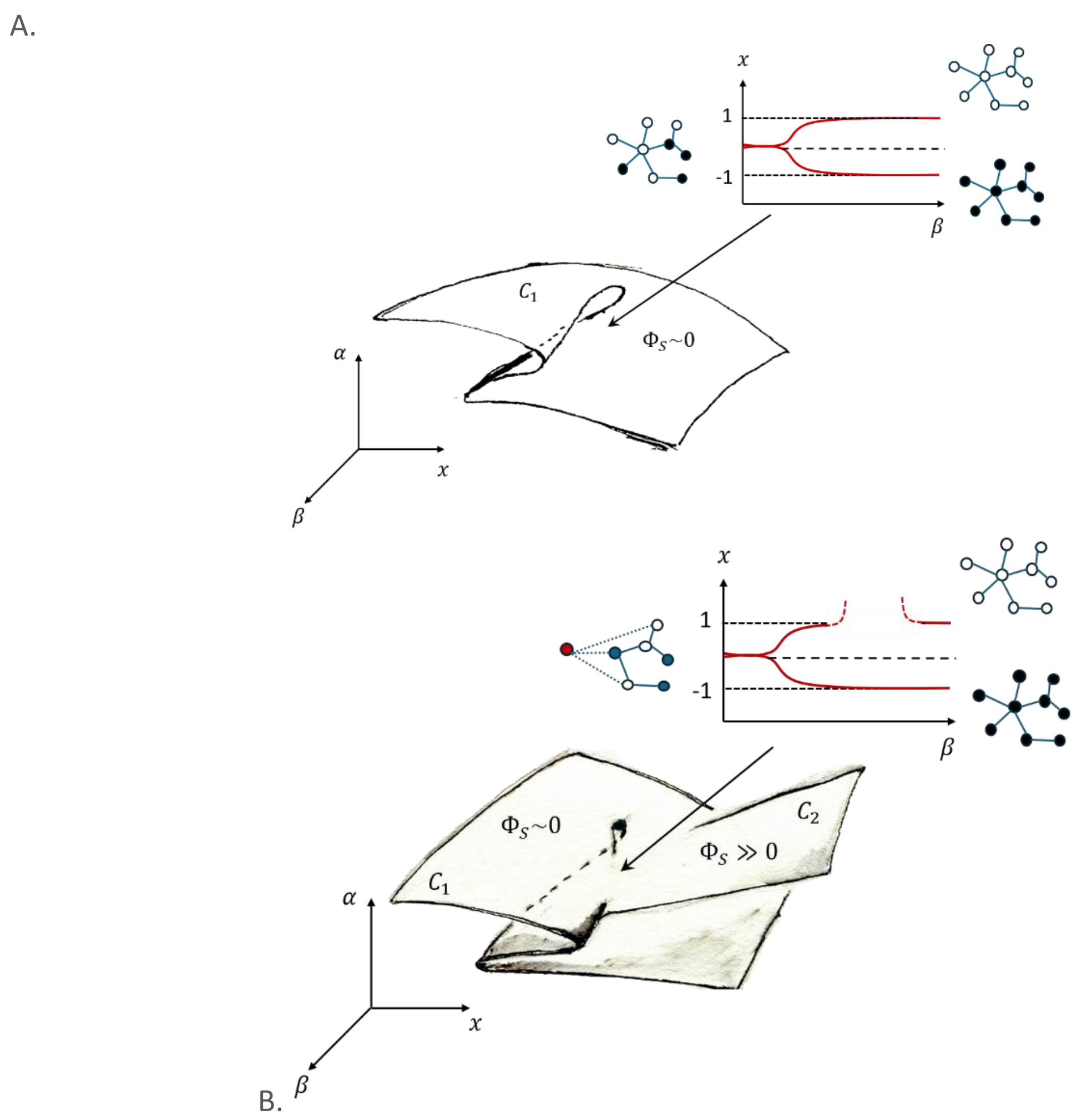

*Figure 6 Phase space of the modified Ising model according to Tom's catastrophe theory*

In the second case, we ask which is the effect of considering an additional context (entangled to the system's states), as discussed in the previous section. In such cases, not only does the topology of the network change in response to different contexts but also the manifold of the phase space, which is related to a large $\Phi_S$ entropy. In other words, the increase in this kind of entropy is related to the rupture of the manifold where the system's dynamics evolve (Figure 6B).

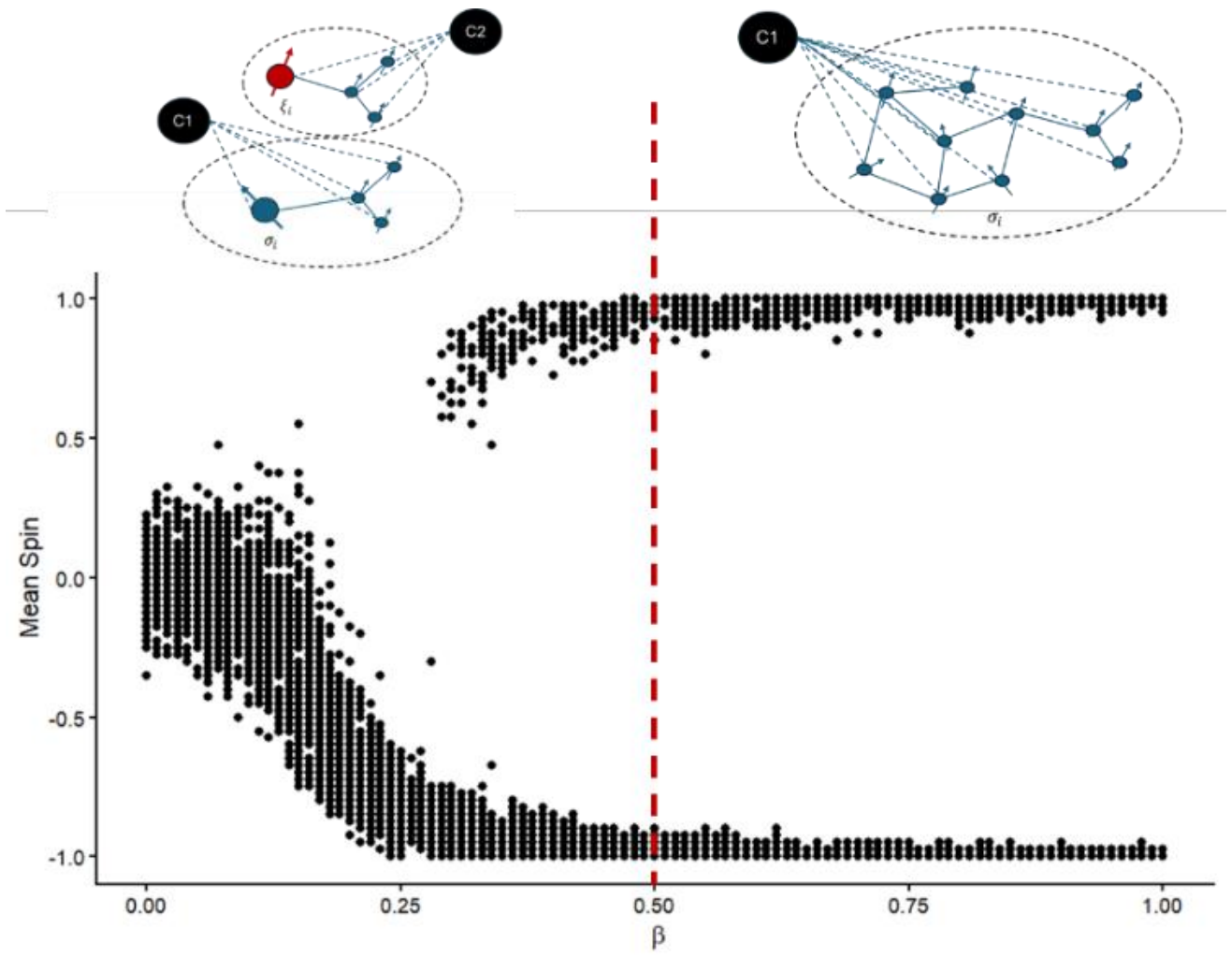

*Figure 7 Visualization of conceptual background Ising networks implemented in this example*

.

On the basis of this theoretical foundation, we implemented a basic Ising model in a random network (without periodic boundary conditions) [29]; the change in element identity was implemented considering a modification of the system's topology when the elements changed their identity. Since it is essentially impossible to keep mathematical account of the second context, we represent such dynamics only from the $C_1$ perspective (see the sketch of how the system's topology changes in Figure 7).

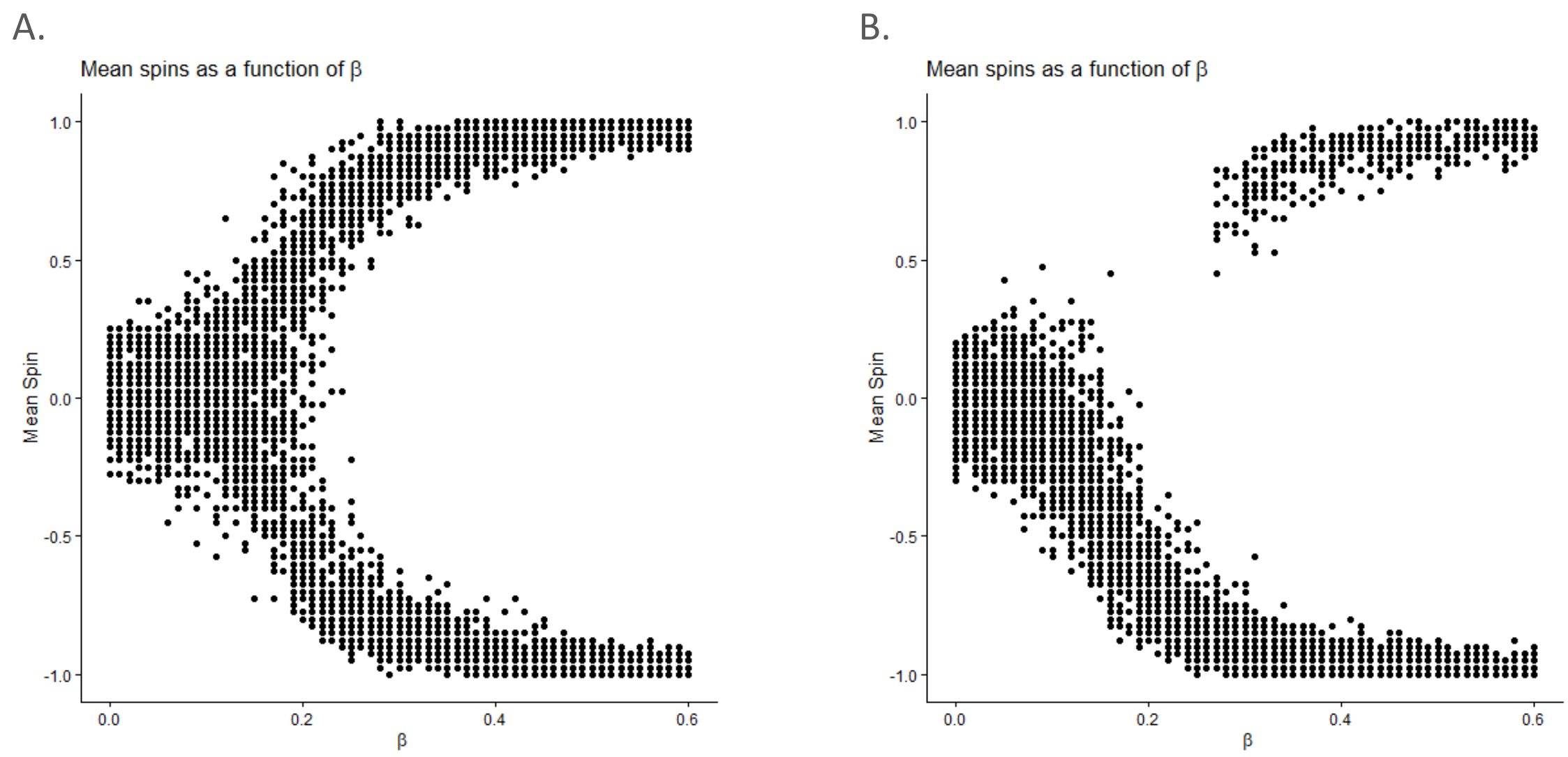

*Figure 8 Phase diagram of a classical Ising model (A) vs. a modified Ising model (B), with a changing inherent topology*

The conventional phase transition and splitting of the phase distribution as a function of the parameter $\beta$ are shown in Figure 8A, which corresponds to the classical manifold of the phase space of a spin network with a critical state, a symmetry rupture and a corresponding phase transition (Figure 6A). These results demonstrate the existence of critical points and phase separation in the random network.

On the other hand, in Figure 8B, we observe the effect of the modified manifold because of the change in the system's structure as a function of the context, where there is a partial symmetry rupture and a nonlineal transition once the system recovers the original system's structure. Accordingly, $\boldsymbol{\Phi_S} = 0.31 >$ 0 (symmetric result $\boldsymbol{\Phi_S} = 0.22 > 0$), which reflects the divergence of the trajectories and, specifically, the divergence of the intrinsic system's structure. This result thus illustrates how complex systems can have both continuous phase transitions and explosive behaviour (which has been observed in several complex systems)[5] rooted in radical changes in the background system's identity and the modification of the system's manifold.

# Discussion

A priori theoretical and mathematical concepts permeate several definitions and concepts in the natural sciences and are essential for the discovery of universal principles of nature from physics to biology. For instance, Darwin's intrinsic motivation was to search for the fundamental laws of life and nature and, in this way, to discover the evolutionary law of species [32]. A priori concepts allow for an objective definition of the observed and modelled system and led to the definition of collective dynamics and emergent phenomena of natural systems. From this perspective, the system's elements embedded in the dynamics of a system become the fabric of reality.

However, some scientists have noted limits in such aprioristic definitions: while current ways to track complex systems either as mechanistic or statistical models restrict systems to a system's class $S \in gKW_0$ on the basis of the concepts of point as well as on set theory, such approaches often ignore how deep elements are related to their context [33]. In this way, mathematical models consider cross-scales as aggregation (including concepts for higher-order interactions [5]) but ignore potential cross interrelations between elements and their context.

In this study, I considered the entanglement between system parts to represent complex interactions with their context, which led to elements that cannot be trivially reduced to a concept of point (and thus set theory); from this concept, it is possible to induce the notion of elastic states, a concept that has long been suggested in the literature [15,31,34] but which has not been fully formalized.

The formal definition of different forms of entanglement provides a mathematical framework; however, it is an incomplete recording of such complex relationships (from the nature of the system, it remains incomplete). All the different examples applied to graphs not only visualize the graph interpretation of different $S \in gKW_i$ groups but also provide potential mathematical limitations.

To illustrate these principles, I developed a modified Ising model as a $S \in gKW_2$ group that shows a rupture of the phase space and a limitation in the recognition of a phase transition and a critical state. We also followed the concept of "Kantian-Wholes" and introduced for the first time a conceptual framework providing a more formal implementation for this concept [33].

This theory and mathematical concepts beyond the concept of points require different concepts for data and data models. The conjecture about the existence of $gKWs$ could serve as a reminder that only $gKW_0$ can be objectively described and that any data-recording of such elements is invariant and thus objective. In the case of the other $gKWs$, the context is involved and cannot necessarily be recorded absolutely in any form of data.

Such a case should remind us that data are not the truth and that all the data recorded from such elements depend on the intention behind the data collection and subsequent statistical evaluation.

Thus, data are not always the objective and neutral numerical mapping of the properties of systems and elements into a numeric system [35], and systems belonging to $gKW_2$ are a good example of abrupt changes associated with the system's context that lead to problems with data collection and deep problems in statistics that are not limited to epistemic or allosteric uncertainty [36]. Only for $gKW_0$ is this kind of mapping feasible; for the other gKWs, such mapping is challenging and not neutral and is dependent on the subjective perspective of the observer. Thus, not only the statistics but also the data recorded are dependent on the observer (and the observed system).

In case complex systems and living beings belong to $gKW_{i>0}$, this fact could challenge current assumptions about how data and data-based models (such as machine learning models) can lead to mechanistic and reductionistic representations, including assumptions of living beings behaving as computers or machines. Ignoring how context is persistently interlinked in complex elements and systems and how this set fundamental hard limits in control theory has not only scientific but also technical as well societal consequences[1], including concepts such as "programmable" cells or organisms: control and programming work once the context is limited; otherwise, programmable organisms imply the creation of controlled environments as well, where $gKW_{i>0}$ (like humans) are mapped into $gKW_{i=0}$.

This reminds us that, as G. t'Hooft recently affirmed, if we employ statistics in system modelling, then it implies that we do not truly understand the foundations of the system [25], and this is particularly true when the context involved in $gKW_i$ is not available.

The implication of the theory presented in this work is that the existence or nonexistence of something is not taken for granted but is constantly negotiated. Similarly, the relationship between system elements/objects and their environment should not be defined in terms of dualism but rather as dialectics, that is, as a constant dialogue and synthesis between the parts and the whole, a concept that was previously proposed to create completeness in an incomplete (multiscale) universe [1]. Such a concept can also be considered a second form of causality, in which the system defines its constraints through semantics and thus the trajectory in which the energy landscape can be optimized [34].

## Conclusions

Why can exact equations for complex systems not be written? [37]. Mathematical definitions start with a priori concepts, where low-complexity elements can be clearly separated from their context, sampled, and translated into basic and more abstract definitions that lead to set theory and advanced definitions of sets, groups, and spaces. However, complex (biological) systems cannot ultimately be defined a priori, since the intrinsic multiscale and its connection to many potentially plausible contexts must be taken into account, which limits mathematical representation.

---

[1] *The "crooked wood"man must be technically tuned and efficiently straightened. In this way of thinking, the human being is merely biological material; his brain is only a kind of "meat computer"; the human being should become smoothly malleable and optimizable, but above all programmable,* i.e.*, above all do not think for himself, but figure as a technical data body.* https://www.republik.ch/2026/05/09/willkommen-in-der-welt-der-dunklen-aufklaerer

I introduced a taxonomy and a group system to encode different forms of element–environment entanglement, defined in this context as elastic states, and their implementation in the modelling of complex systems. In certain systems, a dialect is expected to provide the correct mathematical basis for a theoretical representation of multiple complex systems rather than a priori definitions.

All the groups $gKW_i$ belong to either classic or quantum-mechanic systems; however, $gKW_2$ is less explored, especially because it represents a mathematical limit; to illustrate how a boundary case could work, we introduced an Ising model that behaves as $gKW_2$ in the boundary case, with nonlinear changes in causal structure (but low variability), which is a complementary case to previous studies of elastic states [31].

This study helps clarify the concept of elastic states and provides a formalism for the concept of Kantian wholes. Despite the intrinsic mathematical constraints they represent, they can provide a basis to better represent the extreme nonlinear behavior of complex systems mathematically and to define a method for setting factual limits in mathematical representation, data-based modelling, and the controllability of complex systems.

## Acknowledgements

The author is thankful to Johann Krocza for the first discussion of the first draft of this manuscript.

## Declaration, use of AI

All the ideas, texts, and figures were generated by the author. No artificial intelligence tool was used to generate substantive content, and only language polishing was supported by the use of such a tool. All the codes were implemented by the author without the assistance of AI tools. The Kullback–Leibre distance was implemented using generative AI. In accordance with the STM taxonomy[2], the artificial intelligence tool used for the preparation of this manuscript belongs to category 1: Language refinement was carried out via Rubriq[3] to detect spelling and typographical errors.

[2] https://s3.eu-west-2.amazonaws.com/stm.offloadmedia/wp-content/uploads/2025/04/23020709/STM_AI_Classification_Recs_19_Sept2025-1.pdf
[3] https://secure.rubriq.aje.com/en/rubriq/editing